# Injury risk increases minimally over a large range of changes in activity level in children

Chinchin Wang[1,2], Tyrel Stokes[3], Jorge Trejo Vargas[3], Russell Steele[3], Niels Wedderkopp[4], Ian Shrier[1]

[1] Centre for Clinical Epidemiology, Lady Davis Institute, Jewish General Hospital, McGill University, 3755 Côte Ste-Catherine Road, Montreal, Quebec, Canada H3T 1E2

[2] Department of Epidemiology, Biostatistics and Occupational Health, McGill University, 1020 Pine Avenue West, Montreal, Quebec, Canada H3A 1A2

[3] Department of Mathematics and Statistics, McGill University, 805 Sherbrooke Street West, Montreal, Quebec, Canada H3A 0B9

[4] Orthopedic Department University Hospital of South West Denmark, Department of Regional Health Research, University of Southern Denmark

**Corresponding Author:**

Ian Shrier MD, PhD

Centre for Clinical Epidemiology, Lady Davis Institute, Jewish General Hospital, McGill University, 3755 Côte Ste-Catherine Road, Montreal, Quebec, Canada H3T 1E2

Email: ian.shrier@mcgill.ca

Phone Number: 1-514-229-0114

## CONFLICTS OF INTEREST

None declared.

## SOURCE OF FUNDING

CW is funded by a Canada Graduate Scholarship and a Project Grant from the Canadian Institutes of Health Research. This project was funded by a grant from the Canadian Institutes of Health Research. The CHAMPS-study DK received funding from the TRYG Foundation, University College Lillebaelt, University of Southern Denmark, The Nordea Foundation, The IMK foundation, The Region of Southern Denmark, The Egmont Foundation, The A.J. Andersen Foundation, The Danish Rheumatism Association, Østifternes Foundation, Brd., Hartmann's Foundation, TEAM Denmark, The Danish Chiropractor Foundation, and The Nordic Institute of Chiropractic and Clinical Biomechanics. Funders had no role in the design of the study, collection, analysis, or interpretation of data, writing the report or the decision to submit the manuscript for publication. No authors involved in the manuscript received any payment to produce the manuscript.

## ETHICS APPROVAL

All participation in the data collection was voluntary with the option to withdraw at any time. The study was approved by the Ethics Committee for the region of Southern Denmark (ID S20080047)

## DATA SHARING

The datasets generated and/or analysed during the current study are not publicly available due to legal and ethical restrictions but are available from the CHAMPS Study Steering Committee ([NWedderkopp@health.sdu.dk](NWedderkopp@health.sdu.dk)) on reasonable request.

## ACKNOWLEDGEMENTS

The authors gratefully acknowledge the valuable work of numerous students who assisted with data collection in the CHAMPS-study DK. They also thank the participating children, their parents, and teachers in the schools involved in the project. They are grateful for the cooperation with The Svendborg Project, Sport Study Sydfyn, and the Municipality of Svendborg.

# ABSTRACT

**Background:** Limited research exists on the association between changes in physical activity levels and injury in children.

**Objective:** To assess how well different variations of the acute:chronic workload ratio (ACWR), a measure of change in activity, predict injury in children.

**Methods:** We conducted a prospective cohort study using data from 1670 Danish schoolchildren measured over 5.5 years (2008 to 2014). Coupled 4-week, uncoupled 4-week, and uncoupled 5-week ACWRs were calculated using activity frequency in the past week as the acute load (numerator), and average weekly activity frequency in the past 4 or 5 weeks as the chronic load (denominator). We modelled the relationship between different ACWR variations and injury using generalized linear and generalized additive models, with and without accounting for repeated measures.

**Results:** The prognostic relationship between the ACWR and injury risk was best represented using a generalized additive mixed model for the uncoupled 5-week ACWR. It predicted an injury risk of ~3% for ACWRs between 0.8 (activity level decreased by 20%) and 1.5 (activity level increased by 50%). When activity decreased by more than 20% (ACWR< 0.8), injury risk was lower (minimum of 1.5% at ACWR=0). When activity increased by more than 50% (ACWR > 1.5), injury risk was higher (maximum of 6% at ACWR = 5). Girls were at significantly higher risk of injury than boys.

**Conclusion:** Increases in physical activity in children are associated with much lower injury risks compared to previous results in adults.

**Word Count:** 239

# INTRODUCTION

Although physical activity is crucial for children's development,[1] higher levels of activity are associated with increases in injury risk and related morbidities.[2,3] There are limited research and guidelines on the relative amount by which children can increase activity while minimizing injury.[4]

The relationship between changes in activity and injury has typically been evaluated across various sports in adults using the acute:chronic workload ratio (ACWR),[5] which can be thought of as a prognostic factor for injury.[6] The ACWR is traditionally calculated as activity in the past week (recent activity; acute load) divided by an unweighted average of activity in the past 4 weeks (past activity; chronic load).[7] This calculation is mathematically "coupled", as the numerator is included in the denominator, and has conceptual limitations.[8,9] An "uncoupled" variation of the ACWR has also been proposed, where the numerator is excluded from the denominator.[10] Furthermore, exponentially weighted moving averages (EWMA) variations have been proposed that apply decreasing weights for activity performed further in the past.[11] There is limited research regarding whether any of these variations are more suitable injury predictors than the coupled ACWR.

An International Olympic Committee (IOC) consensus model identified coupled ACWRs between 0.8 and 1.3 as being associated with the lowest injury risk in adults.[12] It suggested higher risks with ACWRs lower than 0.8 or higher than 1.3.[12] A recent review found inconsistent results for injury risk when ACWR < 0.8.[13] Some studies found higher injury risks with lower ACWRs, some studies found no difference, and some studies found lower risks.[13] A recent randomized controlled trial in elite youth footballers found no reduction in injury risk when coaches were provided instructions and software to ensure that ACWRs were maintained between 0.8 and 1.5 versus no instruction.[14] Because training data for the control group were not reported and these were elite athletes, the control group may have also avoided training programs with large changes in weekly activity.

Despite being presented as validated, the IOC model suffers from limitations that threaten its utility even as a predictive tool.[15] More recent studies have addressed some of these limitations, including accounting for repeated measures.[13] However, the majority of studies discretize the ACWR for analyses and use generalized linear models (GLM; e.g. logistic regression).[13] GLMs restrict the type of relationship between the exposure (ACWR) and outcome (injury) to be the same over the entire range of exposure.[16] Generalized additive models (GAM) are more flexible and might better model heterogenous relationships across the range.[16]

Despite its limitations, the ACWR is currently one of few methods to predict the relationship between relative changes in activity and injury. Our objective is to apply different ACWR variations (coupled and uncoupled) and statistical models (GLM and GAM without and with random effects) to a large sample of children and compare the utility of these different ACWR variations as prognostic factors for the relationship between changes in activity and injury in this population.[6] When results are interpreted as predictive and not causal, ACWR-based analyses can help with clinical management predictions, and generate hypotheses that could later be tested with randomized trials or other appropriate methods that control for confounding.[6]

## METHODS

**Data Source**

This was a prospective cohort study nested within the Childhood Health, Activity, and Motor Performance School Study Denmark (CHAMPS-DK) that followed over 1000 schoolchildren for 5.5 years.[17] A natural experiment occurred in Svendborg, Denmark where some schools increased physical education (PE) to six classes per week while others remained at two. CHAMPS-DK evaluated the health outcomes of the children in these different schools and has over 50 published papers. All children in the thirteen primary schools that agreed to participate were eligible for the study.

Our study uses physical activity and injury data (defined as patient-reported musculoskeletal pain) collected via SMS messages from November 2008 to June 2014. Parents were asked each week whether their child experienced musculoskeletal pain in the past week, and whether the pain was new or continuing from a previous injury. Parents were also asked for the number of times their child participated in leisure-time activity in the past week. This was added to the number of school activity sessions to get a total activity frequency for the week. PE classes were 45 minutes each. After accounting for changing and showering, approximately 30 minutes were dedicated to physical activity.[18] Because organized physical activity sessions are typically approximately 60 minutes or longer, we considered students who had 2 PE classes/week to have 1 school-based activity/week, and students who had 6 PE classes/week to have 3 school-based activities/week.

Participants could enter or leave the study at any time. We included all participants who had sufficient data to calculate ACWRs (at least 5 consecutive weeks). Data were collected throughout the school year (September to June). Missing data during school years were multiply imputed using resampling with matching with five datasets.[19] Where ten or more weeks of data were missing in a row for an individual, these weeks were censored and excluded from analyses.

**ACWR Variations**

The "workload" in the ACWR represents a construct of physical stress or strain to tissue that might result in injury.[20] In our study, we considered the number of recreational and school-based physical activity sessions as load. The exposure was one's ACWR for the week, calculated as acute load divided by chronic load. For the coupled 4-week ACWR, acute load was defined as the activity frequency in the index week (week of calculation) and chronic load was defined as the average activity frequency across the index week and previous 3 weeks (numerator included in denominator). For the uncoupled 4- and 5-week ACWRs, acute loads were the activity frequency in the index week and chronic loads were the average activity frequency in the previous 3 and 4 weeks respectively (numerator excluded from denominator).

Conventional EWMA variations of the ACWR use daily loads.[11] We explored a modified EWMA where the acute load was the unweighted index week and chronic load was an unequally weighted average of the previous 3 or 4 weekly loads. These models had much poorer fit than other ACWR variations and were dropped from consideration.

**Outcome Definition**

We defined injury as any athlete-reported musculoskeletal pain in the index week.[21]

**Statistical Analyses**

The relationship between 3 variations of the ACWR (coupled 4-week; uncoupled 4-week; uncoupled 5-week) and injury (dichotomized to yes/no) was modelled using various regression approaches. For each variation, data were analyzed using GLMs and GAMs without random effects, and with a random intercept for individuals (mixed models; GLMM and GAMM). While we assessed the effect of including a random intercept for school, this effect was non-significant (p=0.5) and models were qualitatively very similar. The number of thin-plate spline basis functions was set to 7 for GAM and GAMMs. Each model used a logit link and treated the ACWR as an underlying continuous variable. The Akaike information criterion (AIC) was used to assess goodness of fit.[22] Since comparing AICs requires each model to use the same index weeks for the outcome observations,[22] we restricted the data to index weeks where 5-week ACWRs could be calculated and based AIC comparisons on these models. A detailed description of model selection is provided in the Appendix.

Models were superimposed to compare ACWR variations and model types. To visualize the consequences of including random effects in GLMMs and GAMMs, we included observed probabilities of injury for each value of the uncoupled 4-week ACWR, discretized to the nearest 0.1. We included 95% confidence intervals (95% CI) accounting for random effects and histograms with the number of entries at each discretized ACWR value to illustrate uncertainty.

The uncoupled 5-week GAMM appeared to best predict the relationship between the ACWR and injury risk based on AIC and model comparisons. Therefore, to assess the significance of gender, fixed effects were included in separate GAMMs for the uncoupled 5-week ACWR and their p-value calculated. Sensitivity analyses were conducted excluding those who performed no activity during a given week (ACWR = 0) or those who did not change their activity over consecutive weeks (ACWR = 1) to assess their potential influence.

As a post-hoc sensitivity analysis, we explored whether increasing the number of chronic load weeks from 4 (uncoupled 5-week ACWR) to 5 (uncoupled 6-week ACWR) or 6 (uncoupled 7-week ACWR) improved model fit. We assessed goodness of fit using AICs, restricting the data to index weeks where 7-week ACWRs could be calculated. We based AIC comparisons on these models.

Based on our best prognostic factor model (uncoupled 5-week GAMM), we calculated the sample sizes required to detect a significant effect of doubling activity on injury risk using various randomized trial designs. Analyses were conducted in R 3.6.0, specifically the lme4[23], mgcv[24], and gamm4[25] packages.

# RESULTS

Out of 1755 children who participated in CHAMPS-DK, 1660 children aged 6 to 17 were included in our study and followed for an average of 3.8 years. This represented 286,536 weeks of data, of which 11,458 (4%) had injury. Children with SMS data were generally similar in age, gender and school type as children without SMS data (Table 1). Data on total household income and birthplace were mostly missing in non-participants. Overall, participants engaged in a mean of 1.6 (SD: 1.1) leisure-time activity sessions per week. The most common activities were soccer (played at least once by 66% of participants), swimming (59%), and handball (53%). Over the course of the study, 91% of participants reported injuries at some point during follow-up. A participant flow diagram is shown in Appendix: Figure S3.

## Uncoupled vs. Coupled ACWR

Traditional GLMs for the uncoupled and coupled 4-week ACWRs are presented in Figure 1. The coupled and uncoupled 4-week ACWRs use the same data, but have different denominators. The coupled ACWR is constrained to $\leq 4$.[10] In our data, the coupled ACWR reached a maximum of 3.5, whereas the uncoupled 4-week ACWR extended upwards to ACWR = 21.0 (Figure 1A). Whereas mean injury risk reached 8% for the coupled 4-week ACWR, risk extended to 62% for the uncoupled 4-week ACWR under GLMs (Figure 1B). Although the AIC for the coupled GLM was lower than the uncoupled GLM (coupled: 80,912; uncoupled: 80,959), we focus on the uncoupled ACWR for subsequent analyses because it allows for a wider unrestricted range of ACWR values and better differentiation of exposure patterns.

## Uncoupled 4-week vs. 5-week ACWR

AICs for uncoupled 5-week models were lower than the uncoupled 4-week ACWR model, indicating that inclusion of the extra week in the uncoupled 5-week denominator improved model fit (Table 2). The inclusion of further weeks in the chronic load only slightly improved fit, with an apparent plateau in AICs after 5 weeks (Appendix: Figure S4). As recommended, we focus on the uncoupled 5-week ACWR, representing the simplest model with improved fit for remaining analyses.[22,26]

**Incorporating Random Effects (Mixed Models)**

GLMMs and GAMMs account for repeated measures by including a random intercept for individuals. Compared to models without random effects (GLMs and GAMs), they had consistently better fit across all ACWR variations (Table 2), and predicted a lower injury risk across the uncoupled 5-week ACWR range (Figure 2A and 2B). These results are consistent with the general belief that individuals with different characteristics have different baseline probabilities of getting injured. We focus on mixed models for the uncoupled 5-week ACWR for remaining analyses.

**Generalized Linear vs. Additive Mixed Models**

GLMMs assume that the function describing the relationship between exposure (ACWR) and outcome (injury) is constant across the range of exposure. In contrast, GAMMs allow for multiple functions across the ACWR range. Figure 2A and 2B display the GLMM and GAMM for the uncoupled 5-week ACWR, with a limited range for clarity (full range in Appendix: Figure S5). In our data, the GLMM predicted exponential increases in risk throughout the ACWR range. When ACWR = 1 (activity level unchanged), injury risk was 3%. With ACWR < 1 (decreased activity), the GLMM predicted a gradual decrease in risk to 2% at ACWR = 0 (relative risk compared to ACWR =1; $RR_{ACWR=1} = 0.7$). With ACWR > 1 (increased activity), the GLMM predicted an increase in risk to 4% at ACWR = 2 ($RR_{ACWR=1} = 1.3$), 8% at ACWR = 6 ($RR_{ACWR=1} = 2.7$) and to 24% at the maximum ACWR of 9.3.

In contrast, the GAMM predicted heterogenous relationships across the ACWR range (Figure 2B). There were minimal differences in risk for ACWRs between 0.8 and 1, followed by gradually lower risks with lower ACWRs, reaching 1.5% at ACWR = 0 ($RR_{ACWR=1} = 0.5$). Minimal differences in risk were predicted with ACWRs ranging from 1 to 1.5. For ACWR >1.5, there were gradually higher risks reaching 4% at ACWR = 2 ($RR_{ACWR=1} = 1.3$) and a maximum of 6% at ACWR = 5 ($RR_{ACWR=1} = 2.2$), with large uncertainty at higher ACWRs. Because there were very little data above ACWR > 3 (Figure 2A and B; histograms), confidence intervals should be very wide at high ACWRs. While this was apparent in the GAMMs (Appendix: Figure S5), the GLMMs had unrealistically narrow confidence intervals over the same range.

Because GAMMs had lower AICs than GLMMs, better modeled heterogenous relationships, and had more realistic confidence intervals, we focus on the uncoupled 5-week GAMM for subsequent comparisons.

**Stratification by Gender**

Uncoupled 5-week GAMMs stratified by gender suggested similar relationships between boys and girls at low ACWRs, but that girls may have a higher injury risk than boys at ACWR > 2 (Figure 3). There was a statistically significant association of gender when included in the overall GAMM ($p = 0.047$).

**Sensitivity Analyses**

We examined two potentially influential points in our data. First, 2% of data were from weeks where no activity was performed (ACWR = 0). Although these points were at the extreme of the x-axis with low uncertainty and might have had high leverage, the overall relationship after exclusion was qualitatively similar (Appendix: Figure S6). Second, 30% of data were from weeks where activity was unchanged from the previous 4 weeks (ACWR = 1). GAMMs for the uncoupled 5-week ACWR including and excluding these weeks were also similar in shape (Appendix: Figure S6).

**Sample sizes for randomized trials**

Based on estimates of risk from the previous section, a simple randomized trial examining the effects of doubling activity on injury risk in children (risk = 3% at ACWR = 1 versus risk = 4% at ACWR = 2) would require 11,000 participants (5,500 per group). A repeated cross-over trial with weekly follow-up over 2 years without a washout period for cross-over effects (a typically unreasonable assumption) would require 665 participants (calculations in Appendix).

# DISCUSSION

The predictive relationship between the ACWR and injury risk in our data was best represented by the GAMM for the uncoupled 5-week ACWR. There were minimal differences in risk when activity had increased by up to 50% or had decreased by up to 20% ($0.8 \leq ACWR \leq 1.5$), with risk around 3.0%. Further decreases in activity were associated with lower injury risks to a minimum of 1.5% at ACWR = 0. Larger increases in activity were associated with higher injury risks to a maximum of 6% (2-fold higher risk) at ACWR = 5 (5-fold increase in activity). There was a statistically significant effect of gender in this relationship, with girls at higher risk of injury at ACWR > 1 than boys.

**Differences between ACWR Variations**

Our study found lower injury risks with the uncoupled versus coupled ACWR for ACWR > 1, with models increasingly diverging at coupled ACWR > 2. Although Gabbett et al. previously reported no significant difference in injury risk between the coupled and uncoupled ACWR, our results are consistent with Figure 1D in their paper.[27] While their analyses were based on discretized ACWRs, their figure shows diverging risks when ACWR > 1.99 between the coupled (~21% risk) and uncoupled ACWR (~14% risk). Although Gabbett et al. stated that the differences in injury risk shown in Figure 1D were not significant, they only had a sample size of 28 individuals.

The uncoupled ACWR also has a simpler calculation and interpretation than the coupled ACWR. For the 4-week ACWR, it corresponds directly to the relative increase in activity compared to the previous 3 weeks. Conversely, the coupled ACWR is a proportion of current activity relative to activity in the current and previous 3 weeks. Therefore, the uncoupled ACWR is a more useful measure.

Inclusion of an additional week in the uncoupled ACWR calculation improved model fit, suggesting that activity that occurred 5 weeks previous was associated with injury in our study context. We emphasize that we used a prognostic factor[6] approach to evaluate the relationship between ACWR and injury. Developing and validating an outcome risk score based on prognostic modelling methods[28] would require much more detailed data. Causal interpretations would also require different models and assumptions.

**Linear vs. Additive Models and Random Effects**

Previous studies typically modelled the ACWR-injury risk relationship using logistic regression (i.e. GLM),[29,30] sometimes with quadratic terms.[7,31–34] Like our study, some used mixed models (i.e. GLMM),[31,32,35,36] to account for repeated measures, or generalized estimating equations[37,38]. Many studies did not account for repeated measures.[7,29,30,33]

This is the first study to systematically explore these different models on the same data. In our data on children, mixed models predicted lower injury risks across the ACWR range (Figure 2). These differences demonstrate the importance of accounting for repeated measures to obtain population estimates.

This is also the first study to use GAMMs to model the ACWR-injury risk relationship. GAMMs were able to model heterogenous relationships in the data, including a stable injury risk around ACWR = 1 (Figure 2). GLMMs enforce a single functional form, which is unlikely to be true, and predicted exponential increases across the ACWR range. GAMMs also better accounted for uncertainty at high ACWRs where there were few data. While including polynomial (e.g. quadratic) terms would provide added flexibility over simple log-linear logistic regression,[39] polynomial terms would not be able to model local functional variation in the same way as the GAMMs.[40] For example, the IOC consensus model used a quadratic function that suggested increasing risk below ACWR = 0.8,[33] even though there is no biological explanation for this relationship. Therefore, we believe that GAMMs better predict the ACWR-injury risk relationship.

**Comparison to Other Studies**

Most other studies evaluated the relative risk of injury by comparing discrete categories of ACWR, making it difficult to directly compare results.[13] Similar to many other studies in adults and youth restricted to time-loss injuries, we found an association between higher ACWRs and increased injury risk, although the magnitude of increased risk was smaller.[13] The increase in risk in our model was gradual and only doubled ($RR_{ACWR=1} = 2$) despite a quintupling of activity (Figure 2B). This contrasts with the IOC consensus model, which predicted rapidly increased risk beyond a coupled ACWR of 1.5 (uncoupled ACWR = 1.8).[12] The more gradual and minimal increase in risk in our study may be attributable to differences between adults and children, definitions of load and injury, or other differences in methods.

Our model predicted minimal differences in injury risk when activity increased by up to 50% or decreased by up to 20% (0.8 < uncoupled ACWR < 1.5), similar to the IOC consensus model which suggested a "sweet spot" for coupled ACWRs between 0.8 and 1.3 (0.8 < uncoupled ACWR < 1.4).[12] Other studies have noted increased risk for ACWRs over 1.1.[13] Unlike the IOC consensus model that suggested higher injury risks when activity had decreased by more than 50%, our model suggested lower risks for ACWR < 1. This lowering was still present, albeit smaller, when we excluded data where ACWR = 0. A recent review[13] found that 8/18 analyses (11 papers) that defined load as session ratings of perceived exertions observed the lowest injury risk with lowest ACWR. However, 3/18 observed higher risks at low ACWR, and 7/18 observed no significant differences in risk at low ACWR. Results were qualitatively similar when using other measures of load such as distance, high intensity running, sprinting and accelerations. As injury risk is expected to decrease when there is less exposure time and less forces applied (i.e. lower ACWR), the absence of a decreased risk at low ACWR in the IOC consensus model and some other studies might be due to inappropriately grouping the continuous ACWR into categories prior to analysis,[6] an artifact from using a quadratic function, or particular biases related to measurement error, small sample bias if there were few records and other traditional biases.

**Limitations**

Despite a very large sample, we observed very few weeks at high ACWRs. We calculated ACWRs using 4 or 5 weeks of data as per previous literature. While our post-hoc analyses showed that including additional weeks minimally improved model fit, more research is required to identify the most relevant time windows.[19] Additionally, our load definition encompassed sports of different types and duration. A definition with less variation might increase precision, and our findings may not be generalizable to specific sporting contexts. While we defined injury as any musculoskeletal pain, other definitions (e.g. physician-diagnosed injury; time-loss injury[41]) might necessitate different models. Finally, the ACWR has serious limitations for assessing causality.[15] More advanced methods that account for time-dependent confounding need to be developed to evaluate the causal effect of changes in activity on injury risk in children.

**CONCLUSION**

Using the largest sample size data to date for our objective, the relationship between the ACWR and injury risk in children was best predicted using a GAMM for the uncoupled 5-week ACWR. Injury risk remained around 3% when activity levels had increased by up to 50% or had decreased by up to 20% (0.8 ≤ ACWR ≤ 1.5). Injury risk was lower when activity levels had decreased by more than 20% (ACWR < 0.8) to a minimum of 1.5% at ACWR = 0. Injury risk was higher when activity had increased by more than 50% (ACWR > 1.5) to a maximum of 6% (2-fold higher risk) at ACWR = 5 (5-fold increase in activity), considerably less than reported in adult studies using less flexible methods. Although the ACWR has important limitations, we recommend that when implemented, researchers and practitioners use the uncoupled measure, account for repeated measures, and move beyond logistic models.

# TABLES

Table 1. Baseline characteristics of participants included in study. Characteristics were measured at time of enrollment into the Childhood Health, Activity, and Motor Performance School Study Denmark. Participants were included if they provided data allowing at least one ACWR to be calculated.

|  | Total n=1741 | Included n=1660 | Excluded n=81 |
|---|---|---|---|
| **Gender** | | | |
| Boy | 803 | 769 (46%) | 34 (42%) |
| Girl | 874 | 846 (51%) | 28 (35%) |
| Unknown | 64 | 45 (3%) | 19 (23%) |
| **Grade (age)** | | | |
| 0-1 (6-9) | 673 | 644 (39%) | 29 (36%) |
| 2-3 (8-11) | 741 | 709 (43%) | 32 (40%) |
| 4 (10-12) | 327 | 307 (18%) | 20 (25%) |
| **School type** | | | |
| 3 PE sessions*/week | 995 | 955 (58%) | 40 (51%) |
| 1 PE sessions/week | 746 | 705 (42%) | 41 (49%) |
| **Total Household Income** | | | |
| < kr 400,000 | 179 | 178 (11%) | 1 (1%) |
| kr 400,000 to 599,000 | 351 | 351 (21%) | 0 (0%) |
| kr 600,000 to 799,000 | 345 | 344 (21%) | 1 (1%) |
| > kr 800,000 | 184 | 184 (11%) | 0 (0%) |
| Unknown | 682 | 603 (36%) | 79 (98%) |
| **Birthplace** | | | |
| Denmark | 1290 | 1265 (76%) | 25 (31%) |
| Outside Denmark | 47 | 46 (3%) | 1 (1%) |
| Unknown | 404 | 349 (21%) | 55 (68%) |

\* PE session: Physical education activity sessions where 2 physical education classes counted as 1 session to account for time spent changing and showering

Table 2. Akaike information criteria (AIC) for injury as a function of the uncoupled acute:chronic workload ratio (ACWR) for generalized linear models (GLM), generalized linear mixed models (GLMM), generalized additive models (GAM), and generalized additive mixed models (GAMM). We restricted data to entries where the uncoupled 5-week ACWR could be calculated (i.e. entries preceded by at least 4 weeks of activity data) to directly compare AICs.

|  | Uncoupled 4-week | Uncoupled 5-week |
| --- | --- | --- |
| **Generalized linear model** |  |  |
| GLM (No random effect) | 80,959 | 80,931 |
| GLMM (Random effect) | 77,984 | 77,956 |
| **Generalized additive model** |  |  |
| GAM (No random effect) | 80,919 | 80,891 |
| GAMM (Random effect) | 77,951 | 77,927 |

**FIGURES AND FIGURE LEGENDS**

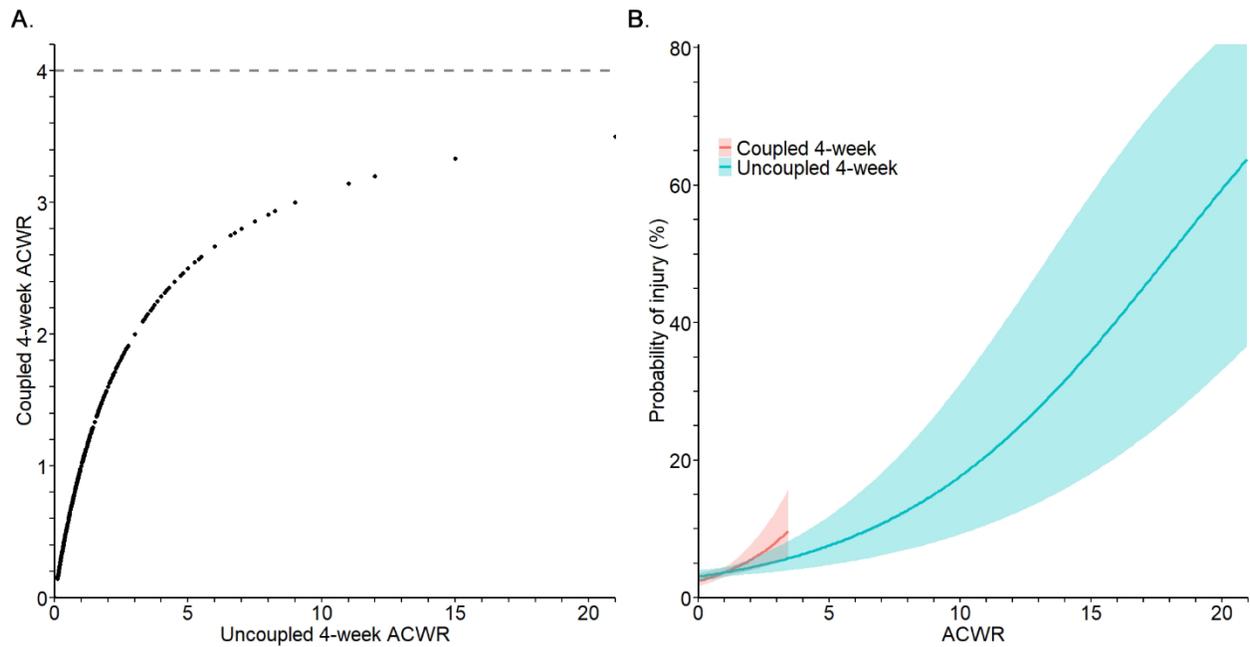

Figure 1. Comparison of the coupled and uncoupled 4-week acute:chronic workload ratio (ACWR) and their relationship with injury in children. A. Relationship between the coupled and uncoupled ACWR. The points correspond to observed values and the dashed line represents the theoretical maximum value of 4 for the coupled ACWR. B. Generalized linear models (logistic regression; no random effect) for the relationship between the coupled and uncoupled ACWR and injury. Lines represent models with 95% CI in shaded areas.

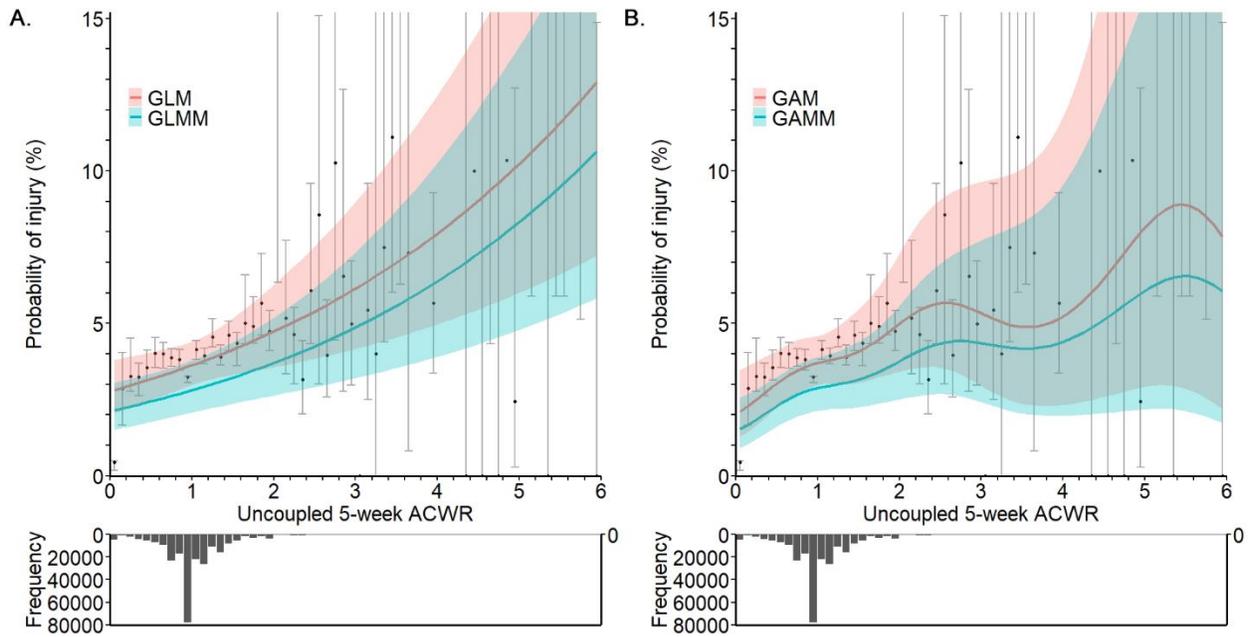

Figure 2. The relationship between the uncoupled 5-week acute:chronic workload ratio (ACWR) and injury in children, without a random effect and with a random effect (mixed model) for individuals. A. Generalized linear model (GLM) and generalized linear mixed model (GLMM). B. Generalized additive model (GAM) and generalized additive mixed model (GAMM). Lines represent models with 95% CI in shaded areas. Points represent observed probability of injury with 95% CI at ACWRs discretized to 0.1. As the point estimates do not account for repeated measures, the GLM and GAM follow the observed points (which also do not account for repeated measures) more closely than the GLMM and GAMM. Histograms show the number of entries at each discretized ACWR. The ACWR range (x-axis) is restricted to ≤ 6 for clarity. The relationship using GLMM and GAMM over the full range of data is illustrated in Figure S3 in the Appendix.

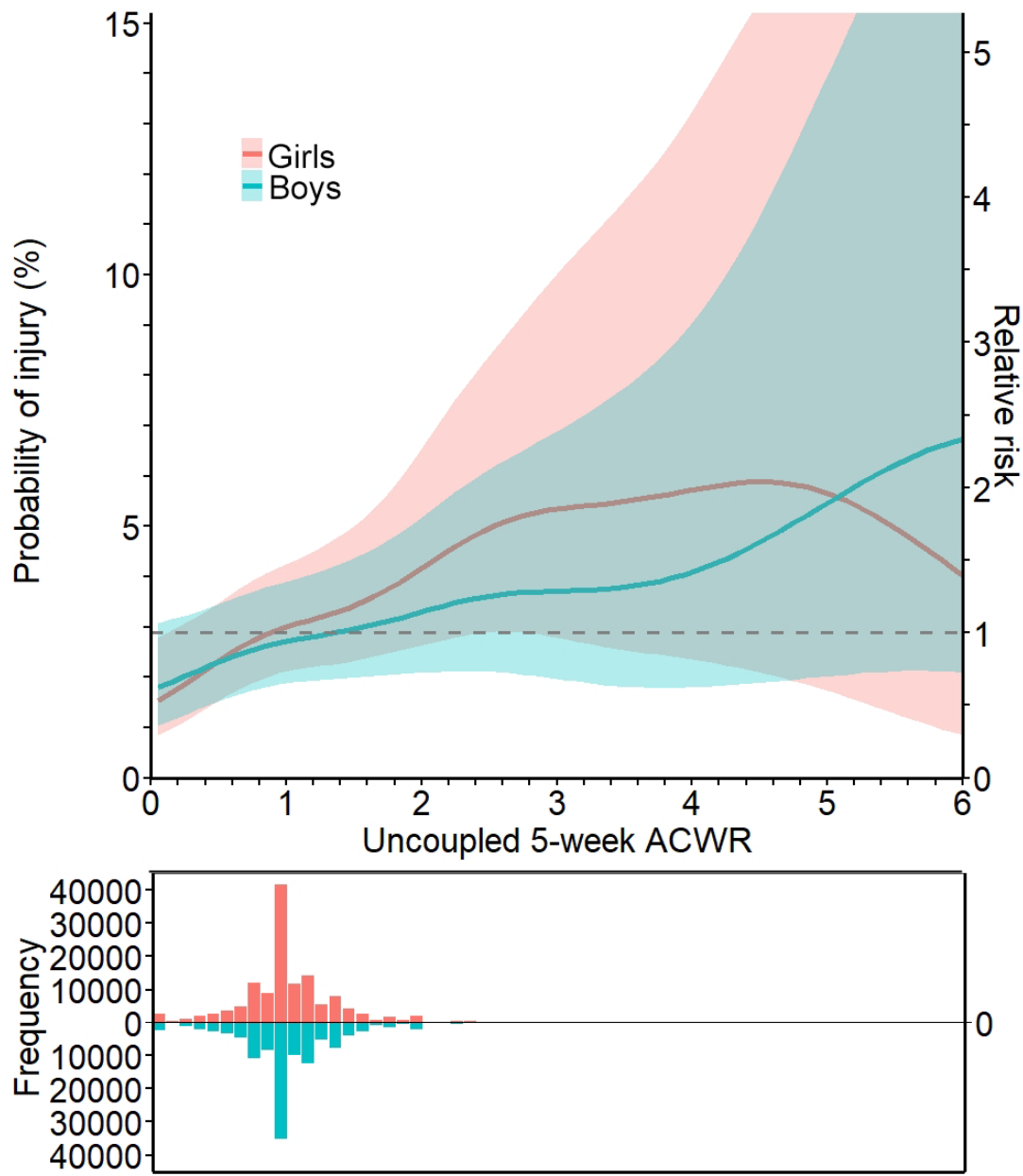

Figure 3. The relationship between the uncoupled 5-week acute:chronic workload ratio (ACWR) and injury in children using generalized additive mixed models with a random effect for individuals; stratified by gender. Lines represent models with 95% CI in shaded areas. Horizontal dashed line represents pooled risk of injury at ACWR = 1, with relative risk versus ACWR = 1 shown on right y-axis. Histogram shows the number of entries at discretized ACWRs by gender. The ACWR range (x-axis) is restricted to ≤ 6 for clarity.

# APPENDIX: Injury risk increases minimally over a large range of the acute-chronic workload ratio in children

## MODEL SELECTION

The default maximum number of basis functions for GAMs and GAMMs using the *mgcv*[1] and *gamm4*[2] R packages is 9. Both R packages then adjust the number of basis functions based on an estimated smoothness parameter. While the estimated smoothness parameter for the GAMs and GAMMs for the uncoupled 5-week and coupled 4-week ACWRs suggested 9 basis functions to model the data, the estimated smoothness parameter for the GAMs and GAMMs for the uncoupled 4-week ACWR suggested 7 basis functions to model the data.

To compare ACWR variations, we set the number of basis functions to be fixed to 7 for all models. Although the uncoupled 5-week GAMM with default 9 basis functions had slightly better fit than the GAMM with 7 basis functions (77,850 vs. 77,927) when averaged over five imputed data sets, this was less than the difference observed between imputed datasets (minimum 77,832; maximum 78,332). The models for each of the 5 imputed datasets is shown in Figure S1. The GAMM with 9 basis functions also appeared to have considerable noise (Figure S2). Therefore, we considered the simpler model with 7 basis functions more appropriate for our analyses.

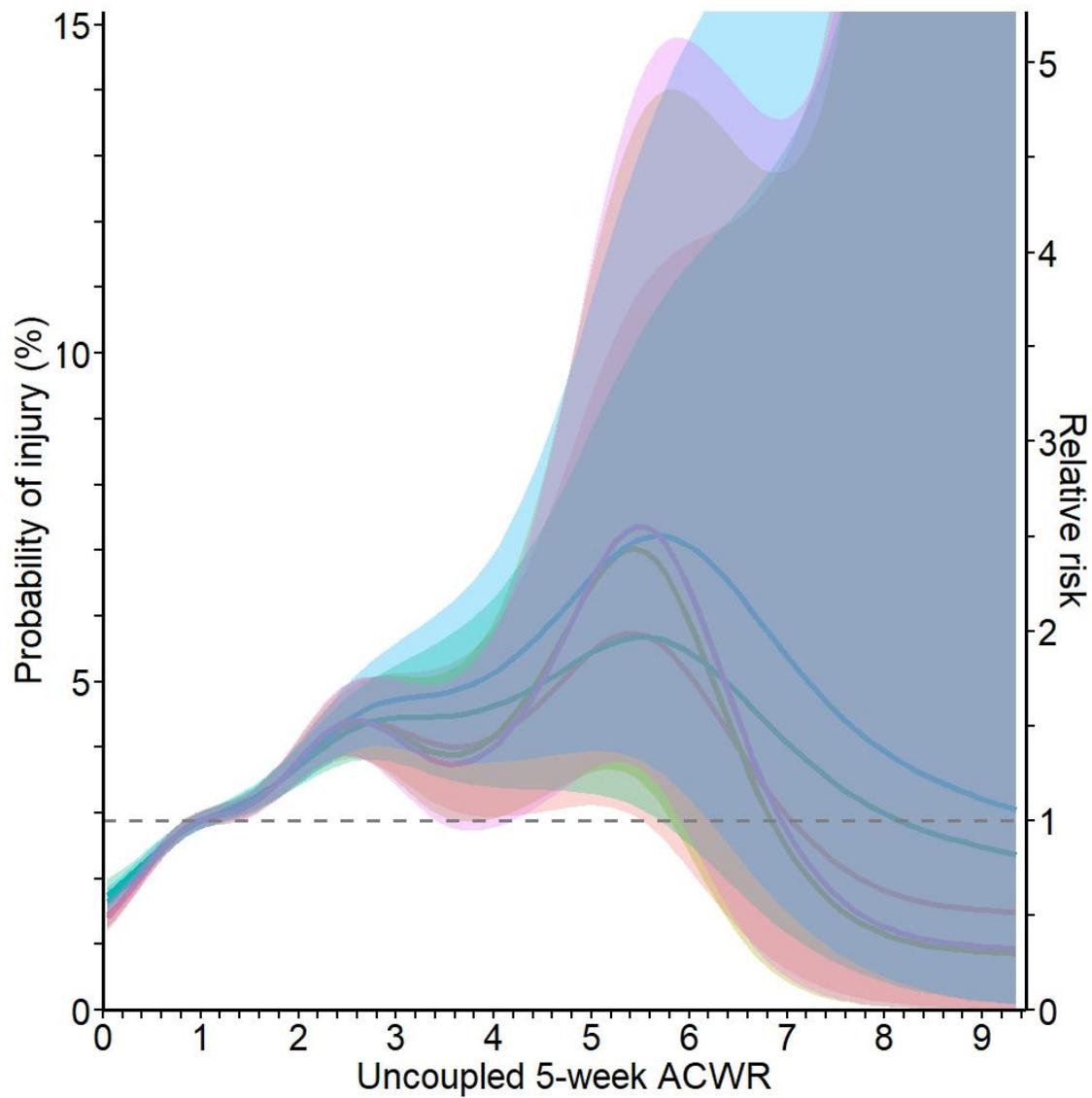

Figure S1. Generalized additive mixed models (GAMM) for the relationship between the uncoupled 5-week acute:chronic workload ratio and injury in children in different imputed datasets. Lines represent models with 95% CI in shaded areas. Akaike Information Criterions for each of the models varied between 77,832 to 78,332.

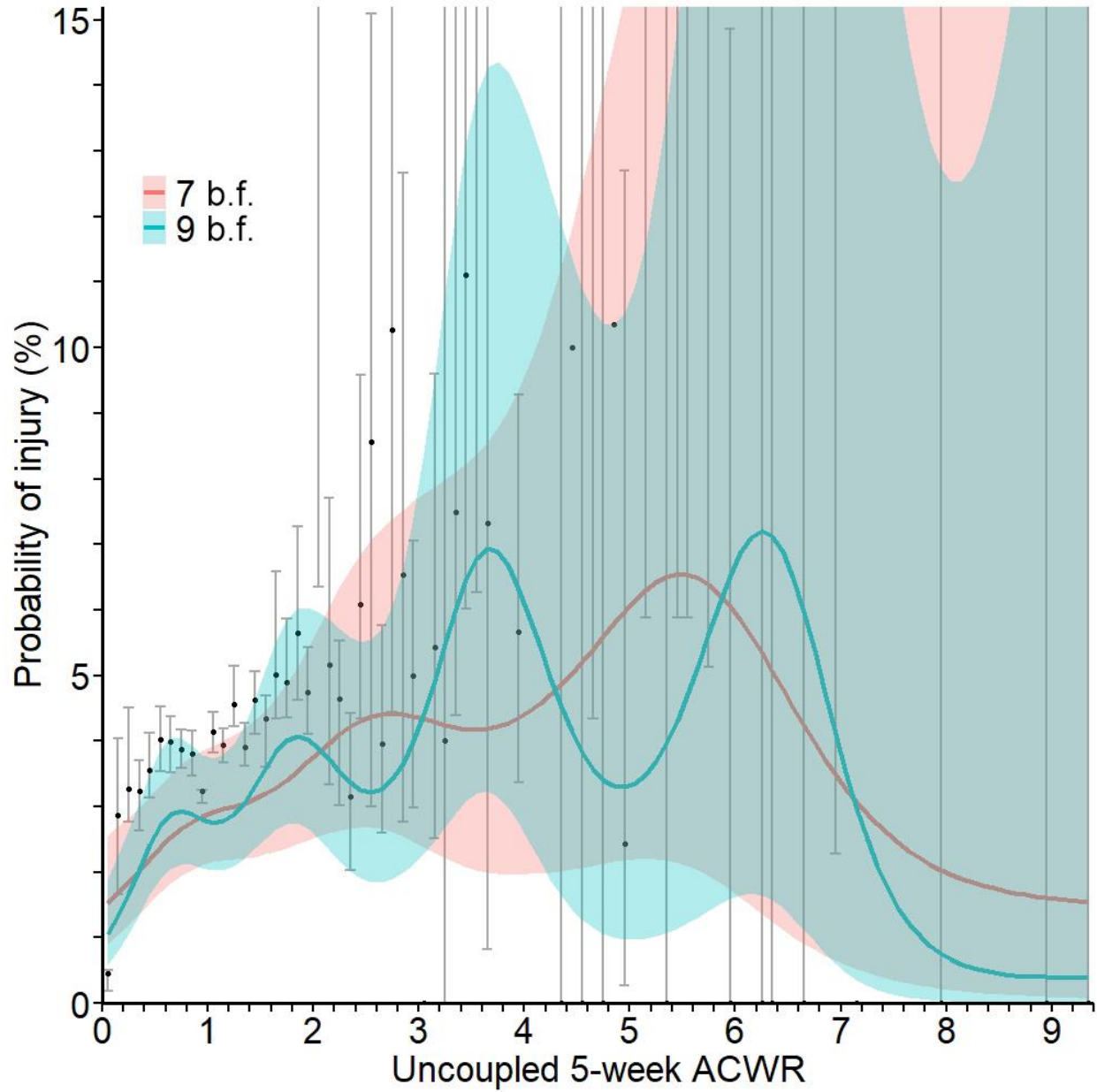

Figure S2. Comparison of generalized additive mixed models (GAMM) for the relationship between the uncoupled 5-week acute:chronic workload ratio and injury in children with 7 basis functions (b.f.) and 9 basis functions. Lines represent models with 95% CI in shaded areas. Points represent observed probability of new onset injury with 95% CI at ACWRs discretized to 0.1.

Table S1. Akaike information criteria (AIC) for new onset pain as a function of exponentially weighted moving average (EWMA) acute:chronic workload ratio (ACWR) variations for generalized linear models (GLM), generalized linear mixed models (GLMM), generalized additive models (GAM), and generalized additive mixed models (GAMM). AICs for unweighted (non-EWMA) models were lower (better fit) and ranged from 77,000 to 84,000.

|  | EWMA ACWR Variation | | |
| --- | --- | --- | --- |
|  | **Coupled 4-week** | **Uncoupled 4-week** | **Uncoupled 5-week** |
| **Generalized linear model** | | | |
| GLM (No random effect) | 96,016 | 90,724 | 90,688 |
| GLMM (Random effect) | 92,422 | 87,441 | 87,404 |
| **Generalized additive model** | | | |
| GAM (No random effect) | 95,840 | 90,457 | 90,438 |
| GAMM (Random effect) | 92,310 | 87,277 | 87,255 |

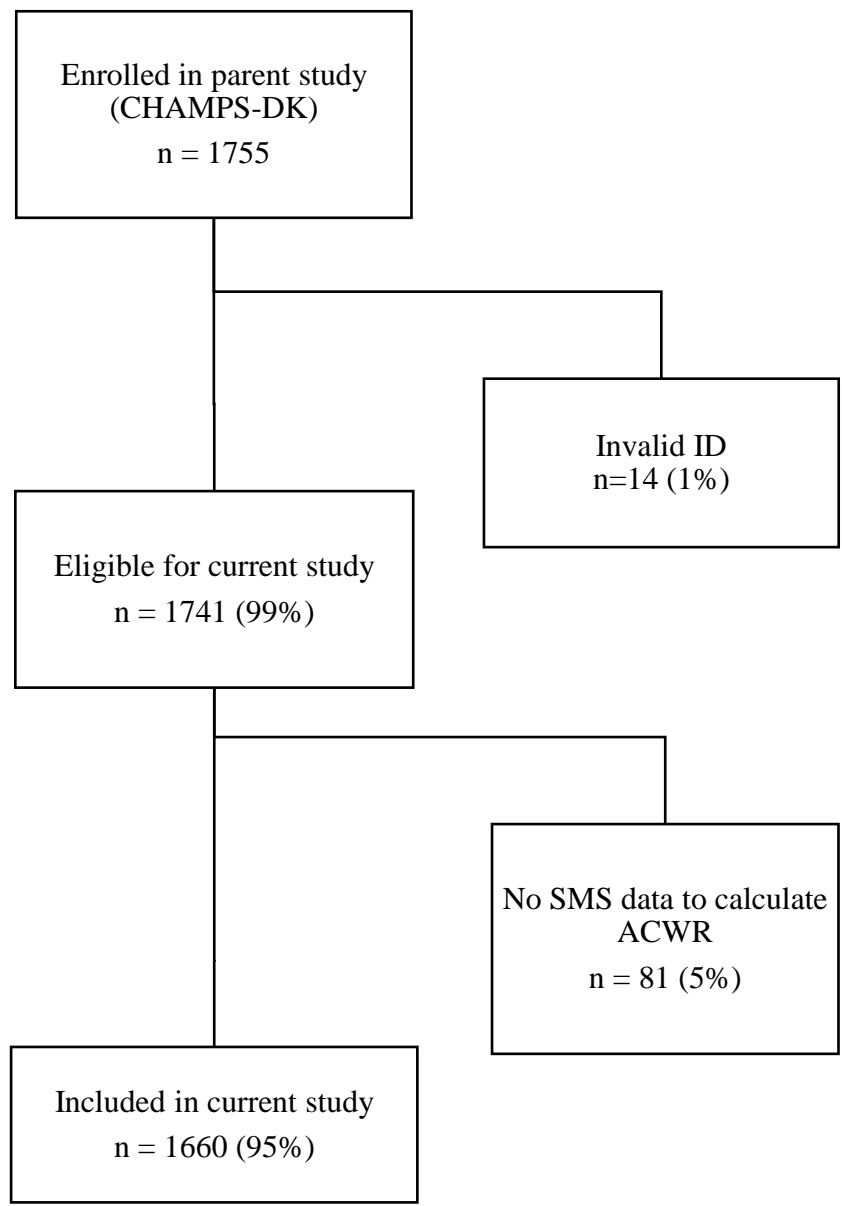

Figure S3. Participant flow diagram for study inclusion. Fourteen individuals were excluded as their ID numbers could not be linked to other study data. Eighty one individuals were excluded because they either did not provide SMS data, or had insufficient data to calculate acute:chronic workload ratios (ACWRs).

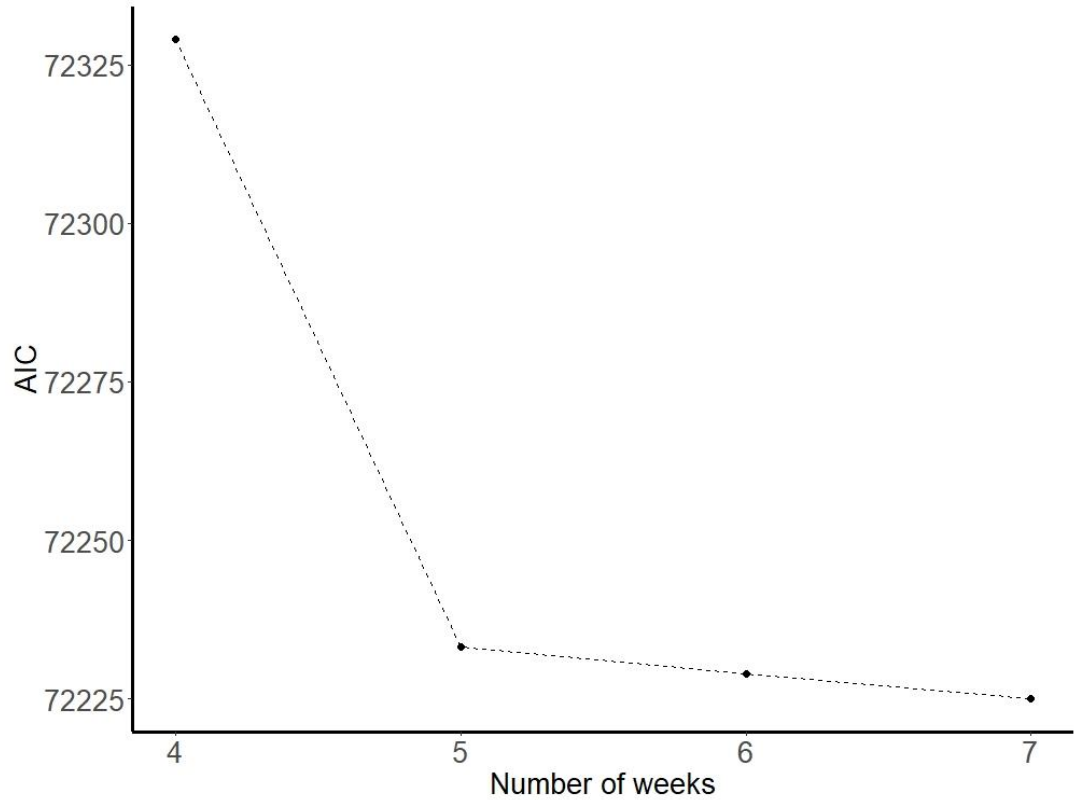

Figure S4. Akaike information criteria (AIC) for injury as a function of the uncoupled acute:chronic workload ratio (ACWR) by number of total weeks using generalized additive mixed models (GAMM). Since comparing AICs requires each model to use the same index weeks for the outcome observations, we restricted data to entries where the uncoupled 7-week ACWR could be calculated (i.e. entries preceded by at least 6 weeks of activity data) to directly compare AICs.

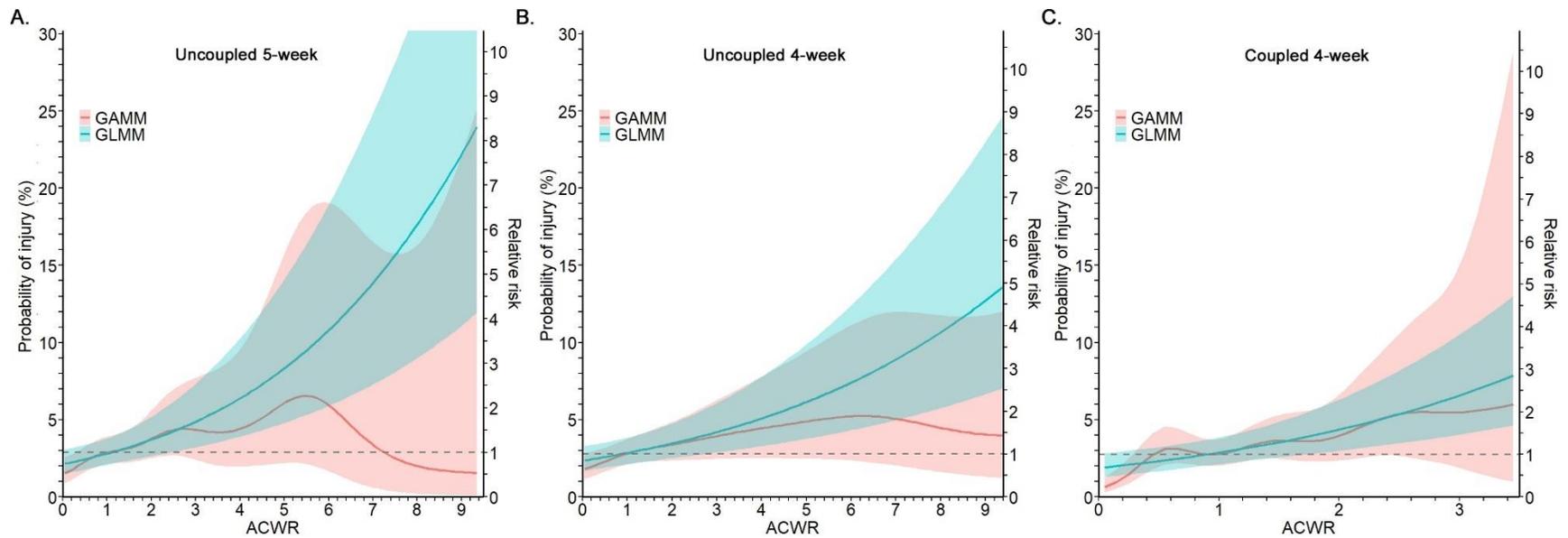

Figure S5. Generalized linear mixed models (GLMM) and generalized additive mixed models (GAMM) for the relationship between variations of the acute:chronic workload ratio (ACWR) and injury in children. A. Uncoupled 5-week ACWR. B. Uncoupled 4-week ACWR. C. Coupled 4-week ACWR. Lines represent models with 95% CI in shaded areas. The full ACWR range (x-axis) is shown for the uncoupled 5-week (maximum 9.3) and coupled 4-week ACWR (maximum 3.4); the range is restricted to ≤ 9.3 for the uncoupled 4-week ACWR for clarity and comparison to the uncoupled 5-week model

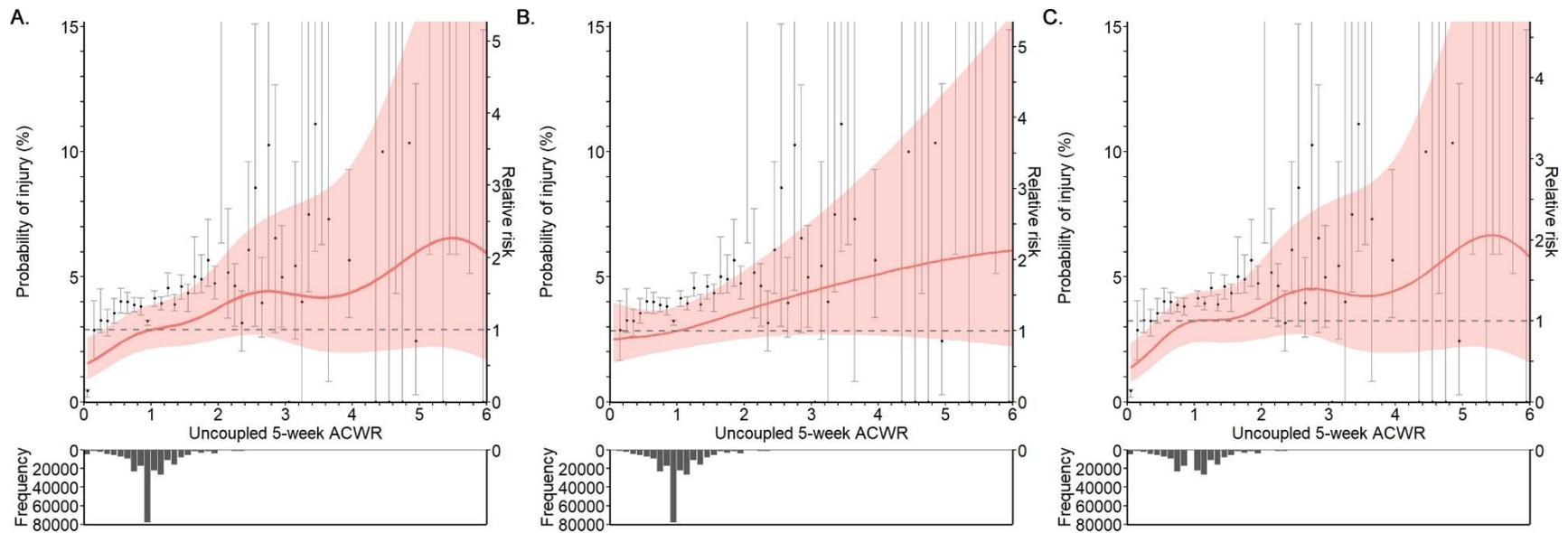

Figure S6. Effect of excluding weeks with no change in activity on the relationship between the uncoupled 5-week acute:chronic workload ratio (ACWR) and new onset pain in children using generalized additive mixed models with a random effect for individuals. A. Including ACWR = 1 (constant activity performed in index and previous 4 weeks). B. Excluding ACWR = 1. Lines represent models with 95% CI in shaded areas. Points represent observed probability of new onset pain with 95% CI at ACWRs discretized to 0.1, not accounting for repeated measures. Horizontal dashed line represents risk of pain at ACWR = 1, with relative risk versus ACWR =1 on right y-axis. Histograms show the number of entries at discretized ACWRs. The ACWR range (x-axis) is restricted to ≤ 6 for clarity.

# SAMPLE SIZE FOR RANDOMIZED TRIAL

**Simple Randomized Trial (5 weeks total)**

Consider a simple randomized trial where participants engaged in the same amount of activity for 4 weeks, followed by 50% of the participants performing one additional week where 1) ACWR = 1 (no change in activity) or 2) ACWR = 2 (doubling activity). To estimate the sample size for such a trial, we used the following numbers:

- alpha = 0.05
- power = 0.8
- injury risk = 3% at ACWR = 1
- injury risk = 4% at ACWR = 2

The calculated sample size is 5,500 per group, or 11,000 participants total. Calculations were performed using PS: Power and Sample Size Calculation.[1]

**Crossover Trial (1 crossover; 10 weeks total)**

It is also possible to conduct a crossover trial where individuals switch from ACWR = 1 to ACWR = 2, or vice versa. Both interventions in the simple randomized trial above require 5 weeks, with the only difference between groups being the activity performed in the last week. For a cross-over trial, for simplicity, we will make the unrealistic assumption that there is no carry-over effect (no need for a washout period after the fifth week where activity either remains unchanged or is doubled).

Let us consider that each participant will perform only one cycle at ACWR = 1 and one cycle at ACWR = 2. Using a correlation between outcomes of 0.17 (as was seen in our data), we would require 2,055 participants. Calculations were performed using PS: Power and Sample Size Calculation.[1]

**Crossover Trial (20 crossovers; 2 years total)**

The required sample size for a crossover trial where each participant performed each intervention many times can be estimated using a method typically used for cluster randomized trials.[2] In this study, a cluster is defined by the repeated cycles of intervention for each participant. Since each intervention is 5 weeks long, we would need to follow participants for almost 2 years to obtain 10 cycles under each condition (cluster size = 20).

The required sample size for the cross-over trial is equal to the sample size for the simple RCT divided by an "inflation factor". The inflation factor is dependent on the intra-class correlation (ICC) within a cluster. In our data, the ICC was 0.11 when restricted to participants with ACWR = 1 or ACWR = 2.

The inflation factor is calculated as:

$$1 + (n - 1) * ICC,$$

where n is the size of the cluster (40 in this hypothetical study). Therefore,

$$Inflation\ factor = 1 + (20 - 1) * 0.11 = 3.09.$$

The required sample size for a cross-over trial with 20 measures per cluster is:

$$\frac{Sample\ size\ not\ including\ Design\ Effect}{Inflation\ Factor} = \frac{2055}{3.09} = 665$$

Therefore, even if we assumed no washout period was necessary and other unlikely assumptions (e.g. no effect of injury on future risk of injury), we would need to follow 665 participants over 2 years to observe a 33% increase in risk (from 3% to 4%).